\documentclass{eptcs}

\providecommand{\event}{PrePost 2016}
\usepackage{graphicx}
\usepackage{listings}
\usepackage{color}

\newcommand{\eat}[1]{ }

\newcommand{\trans}[1]{\stackrel{ #1 }{\longrightarrow}}

\title{Monitoring Assumptions in Assume-Guarantee Contracts\thanks{This work is supported in part by DARPA BRASS program under contract FA8750-16-C-0007 and by ONR SBIR contract N00014-15-C-0126.}}
\author{Oleg Sokolsky \quad Teng Zhang \quad Insup Lee
\institute{University of Pennsylvania\\Philadelphia, PA, USA}
\and
Michael McDougall
\institute{GrammaTech, Inc.\\Ithaca, NY, USA}
}
\def\titlerunning{Monitoring Assumptions in Assume-Guarantee Contracts}
\def\authorrunning{O. Sokolsky, T. Zhang, I. Lee, and M. McDougall}

\date{}

\begin{document}

\maketitle

\begin{abstract}
Pre-deployment verification of software components with respect to behavioral specifications in the assume-guarantee form does not, in general, guarantee absence of errors at run time.  This is because assumptions about the environment cannot be discharged until the environment is fixed.  An intuitive approach is to complement pre-deployment verification of guarantees, up to the assumptions, with post-deployment monitoring of environment behavior to check that the assumptions are satisfied at run time.  Such a monitor is typically implemented by instrumenting the application code of the component.  An additional challenge for the monitoring step is that environment behaviors are typically obtained through an I/O library, which may alter the component's view of the input format.  This transformation requires us to introduce a second pre-deployment verification step to ensure that alarms raised by the monitor would indeed correspond to violations of the environment assumptions.  In this paper, we describe an approach for constructing monitors and verifying them against the component assumption.  We also discuss limitations of instrumentation-based monitoring and potential ways to overcome it.
\end{abstract}

\section{Introduction}
\label{sec:intro}

Behavioral specifications of software components are often written as assume-guarantee contracts $A \Rightarrow G$, where the assumption $A$ describes constraints on acceptable behaviors of the environment and the guarantee $G$ relates well-formed environment behaviors to component behaviors (as in, e.g.,~\cite{Benveniste2008}).  This approach allows us to develop components without complete knowledge of their environment.  However, in an open environment, error-free execution cannot be fully assured before deployment. While the guarantee can be verified at design time, the assumption describes the property of the environment and thus has to be checked at run time.
A natural approach is to complement pre-deployment verification of the contract with post-deployment monitoring of the assumption.  The combination of contract verification and monitoring, effectively, turns component specification into $A \Rightarrow G \wedge \neg A \Rightarrow Alarm$.

Our approach to monitoring is based on instrumenting the application code of the component.\footnote{Since monitoring is from the system perspective, we refer to environment observations as inputs.}  We expect that the application code makes use of some I/O library to communicate with its environment.  We assume that the library supports an API with well-known semantics of each API call.  At the same time, we assume that the implementation of the I/O library is a black box, and therefore we can apply instrumentation in the application code, but not in the library code.  As a result, we do not monitor the assumption directly, but mediated by the I/O API.  We refer to the property being monitored on the application code as the \textit{internal assumption} of the component.  This separation between the internal assumption and the contract assumption brings up the question whether the monitor, constructed from the application code perspective, will correctly detect deviations from the assumption.  Answering this question necessitates the second pre-deployment verification step in our approach.  At the very least, the monitor should not raise any alarms when inputs satisfy the assumption.  

The contribution of this paper is a way for constructing internal assumption monitors from the source code of the component application code and verifying their correctness with respect to the contract assumption.  On the other hand, we show that not all deviations from the format can be detected by monitoring the internal assumption, and that the component may be robust to some deviations.  A precise characterization of limitations of such monitors is left for future work.

We specify contract assumptions as extended finite state machines (EFSM) that reflect temporal relationships between inputs, as well as semantic constraints on inputs, captured as predicates on current and past values of inputs.
We construct the monitors based on the framework for checking compatibility between producers and consumers, developed by Evan Driscoll in his doctoral work~\cite{Driscoll-phd,Driscoll11}.  His approach is based on modeling input and output data formats as visibly pushdown automata~\cite{AM04} enhanced with semantic constraints.  The automata, and to some extent semantic constraints, can be extracted directly from the application code using static analysis techniques.  

In Driscoll's approach, conformance is based on checking language inclusion between models of producer and consumer data formats.  We use a slight extension of this approach to verify conformance between the assumption in the contract and the monitor automaton.  We then monitor inputs of the component with respect to the same automaton.

There exists work in the literature on software verification with respect to assume-guarantee contracts, for example,~\cite{Giannakopoulou04}.  We do not consider this aspect of the problem further in this paper.  Instead, we concentrate on the problem of deriving monitors from the application code and exploring their relationship to the assumption of the component.

The rest of the paper is organized as follows.  Section~\ref{sec:case-study} introduces the case study that we use as a running example throughout the paper.  Section~\ref{sec:construction} describes the construction of monitors from the application source code and verification of monitors with respect to the assumption of the contract.  Section~\ref{sec:discussion} offers a discussion of the power of resulting monitors in its relation to the I/O API chosen by the application code developer.

\section{Case study}
\label{sec:case-study}

In this section, we introduce the case study that motivates our approach.  Consider a simple program for calculating travel distance from GPS coordinates.  Input for the calculator is a stream of data points along the path.  Input can be streamed directly from a sensor, from a pre-processor that, e.g., cleans up the data and performs noise reduction, or from a pre-recorded file.

\begin{figure}
\centerline{\includegraphics[width=.7\textwidth]{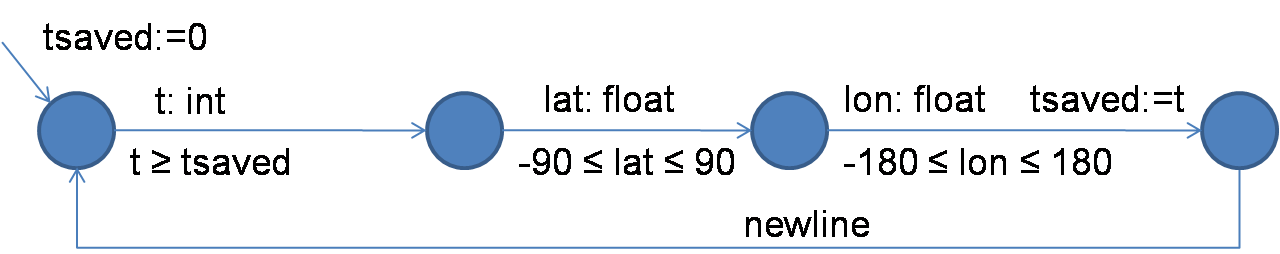}}
\caption{Assumption of the calculator component}
\label{fig:assumption}
\end{figure}

The informal specification of the input format is as follows: 1) input is a sequence of data points; 2) each point is represented as a newline-terminated string; 3) a point consists of three whitespace-separated numbers, which provide timestamp, latitude, and longitude; 4) timestamp is a non-negative integer, latitude is a floating point number in range [-90,90], longitude is a floating point number in range [-180,180]; 5) points are in the non-decreasing order of timestamps.  The specification can be captured by an extended finite state machine (EFSM) shown in Figure~\ref{fig:assumption}.  Note that, for simplicity, we do not represent white space separators in the format specification, but keep the newline terminator to be able to illustrate our point.

Listing~\ref{lst:simple_parser} shows a possible implementation of the calculator (giving only the input-handling code).  It is easy to see that some of the specification is not directly monitorable on the code.  In particular, the parser does not see line breaks, because the \texttt{scanf} function silently consumes white space.  Similarly, the code does not see the the input data as points, but rather as independently read in values.  
The code makes an implicit assumption that values read upon invocations $3i+1$, $3i+2$, and $3i+3$ of the \texttt{scanf} function constitute the $i$th point.  This assumption happens to be satisfied by the input format.  
For comparison, consider a different implementation of the same calculator, shown in Listing~\ref{lst:simple_parser2}.  This version relies on the assumption that each point appears on its own line.  It is also correct with respect to the format specification.  
Each of the two versions is robust to some deviations from the input format specification and vulnerable to others.  We will show in Section~\ref{sec:discussion} that the two versions are vulnerable to very different kinds of deviations. 

\begin{lstlisting}  [float,language = C, frame=single, commentstyle=\color{codegreen}, caption= Simple distance calculator, label = lst:simple_parser]  % Start your code-block

int main(int argc, char * argv[]) {
    int time;
    float lat, lon, last_lat, last_lon;
    float total_dist = 0.0;
    _Bool is_first_pt = 1;

    while (1) {
        int items_read = 0;
        items_read = scanf("%d", &time);
        if (items_read <= 0) break;
        items_read = scanf("%f", &lat);
        if (items_read <= 0) break;
        items_read = scanf("%f", &lon);
        if (items_read <= 0) break;
        ...
    }
    ...
}
\end{lstlisting}

\begin{lstlisting}  [float,language = C, frame=single, commentstyle=\color{codegreen}, caption= Alternative calculator implementation, label = lst:simple_parser2]  % Start your code-block

int main(int argc, char * argv[]) {
    int time;
    char* buffer = NULL;
    size_t bufsize;
    float lat, lon, last_lat, last_lon;
    float total_dist = 0.0;
    _Bool is_first_pt = 1;

    while (1) {
        getline(&buffer, &bufsize, stdin);
        if (items_read == -1) break;
        items_read = sscanf(buffer, "%d%f%f\n", &time, &lat, &lon);
        if (items_read < 3) break;
        free(buffer);
        buffer = NULL;
        ...
    }
    ...
}
\end{lstlisting}

\section{Monitor construction and verification}
\label{sec:construction}

\subsection{Preliminaries and problem statement}

In this section, we define assume guarantee contracts for components in terms of extended finite state machines.

Given a set of typed channels $C$, a \textit{trace} $t$ over $C$ is a sequence $\langle c_1(v_1), c_2(v_2), ... \rangle$, where $c_i \in C$ and $v_i$ is the value transmitted over $c_i$ in step $i$.  Let $\mathbf{T}_C$ be the set of all traces over $C$.  Given a trace $t \in \mathbf{T}_C$, $t \downarrow C' \in \mathbf{T}_{C'}$ is a projection of $t$ on $C'$ defined as a maximal subsequence of $t$, in which every $c_i \in C'$.

Given a set of variables $V$, a \textit{valuation} of $V$ is a mapping that associates a type-correct value to each variable $V$.  Let ${\cal D}^V$ be the set of all valuations for $V$.

A \textit{component} $\cal C$ is a tuple $\langle I, O, P \rangle$, where $I$ is a set of typed input channels, $O$ a set of typed output channels, and $P(I,O)$ is the component logic that consumes data from input channels and computes values to transmit on output channels.  Semantics of a component is given by a set of traces $T^{\cal C} \subseteq \mathbf{T}_{I \cup U}$.

An \textit{extended finite state machine} (EFSM) is a tuple $\langle S, s_0, C, V, D_0, T \rangle$, where $S$ is a set of locations with the designated start location $s_0$, an alphabet $C$ is a set of channels, $V$ is a set of variables, and $D_0$ is the initial valuation of $V$. $T$ is the transition relation.  Each transition, denoted $s_1 \trans{c\{a\}}_g s_2$, is equipped with a guard $g$, which is a predicate over $V \cup \{c\}$ and an update $a$, which is a function ${\cal D}^{V \cup \{c\}}\rightarrow{\cal D}^V$.

A state of EFSM $M = \langle S, s_0, C, V, D_0, T \rangle$ is $\langle s, D \rangle$, where $s \in S$ and $D \in {\cal D}^V$.  Semantics of $M$ is given by a set of runs.  A \textit{run} of $M$ is a sequence $\langle s_0, D_0 \rangle, c_0(v_0), \langle s_1, D_1 \rangle, c_1(v_1), ...$, such that, for each $i$, $s_i \trans{c_i\{a_i\}}_{g_i} s_{i+1} \in T$, $D_i \cup \{c_i \mapsto v_i\} \in g_i$, and $D_{i+1} = a_i(D_i \cup \{c_i \mapsto v_i\})$.  That is, valuation $D_i$ in the source state together with the value on channel $c_i$ satisfy the guard $g_i$, and the action $a_i$ transforms $D_i$ into $D_{i+1}$.  A trace is the projection of a run on the channels, abstracting away the information about locations.  The set of traces of $M$ is denoted $T^M$.

Given a component ${\cal C} = \langle I, O, P \rangle$, an \textit{assume-guarantee contract} $\langle A, G \rangle$ consists of two EFSMs: the assumption EFSM $A$ with the alphabet $I$ and the guarantee EFSM $G$ with the alphabet $I \cup O$.  $\cal C$ satisfies the contract if, for every trace $t \in T^{\cal C}$, $t\downarrow I \in T^A \Rightarrow t \in T^G$.

We model component logic $P(I,O)$ as a composition of three EFSMs: $M_I \otimes M_P \otimes M_O$.\footnote{Intuitively, $\oplus$ is standard product with synchronization over common channels.  We do not define it formally here, since it is never directly computed in our approach.}  $M_P$ represents application code of the component.  Channels in $M_P$ represent values returned by input API calls (denoted as $I'$) and values passed to output API calls (respectively, $O'$).  $M_I$ represents input API calls, transforming environment inputs $I$ into returned inputs $I'$.  $M_O$, similarly, transforms $O'$ into environment outputs $O$.

\paragraph{Problem statement.}
An \textit{internal assumption} of the application code in a component ${\cal C} = \langle I, O, P\rangle$ with the contract $\langle A, G \rangle$ is an EFSM over the set of channels $I'$.  Given EFSMs $M_P$, $M_I$, and $A$, our goal is to construct the internal assumption $M^A_P$ to satisfy the following two requirements: (1) every trace of $M_P$, projected on $I'$, is a trace of $M^A_P$ and (2) every trace of $A$ is a trace of $M_I \otimes M^A_P$. 
We treat EFSM $M^A_P$ as a monitor that raises an alarm whenever a given trace is not a trace of $M^A_P$.
In the next section, we show how to construct $M^A_P$ and $M_I \otimes M^A_P$.

\subsection{Monitor construction and verification}

\paragraph{Monitor construction.}
We use the approach of~\cite{Driscoll11} to extract monitor skeletons --- transitions of the EFSM.  The algorithm given in~\cite{Driscoll11} produces visibly pushdown automata~\cite{AM04}.  However, to simplify presentation, we limit our attention to state machines in this paper, effectively assuming that the application code does not have procedure calls except for external API calls to black-box libraries.

The idea behind the extraction algorithm is to replace calls to I/O APIs with transitions labeled by the type of data item returned by each call.  Following~\cite{Driscoll11}, we limit our attention to C types \texttt{int} and \texttt{float}. We also allow API calls to produce multiple data items.  Many standard C library calls are parameterized by a \textit{format string}.  If multiple data items are produced, we introduce a sequence of transitions to match items in the format string in the order of their appearance.  For example, a call to \texttt{scanf()} with the format string ``\%d\%f'' introduces two transitions: $\trans{int}\trans{float}$.  We refer to this sequence as the \textit{skeleton fragment} of the API call.

The algorithm for constructing the monitor skeleton operates on the control flow graph of the component application code.  If the node in the graph corresponds to a call to an I/O API procedure, the transition from the node is replaced with the skeleton fragment of the API call.  All other transitions become $\epsilon$-transitions and are removed from the skeleton in the standard way.

Once the skeleton is constructed, it is turned into an EFSM of the internal assumption by adding state variables, guards, and updates representing semantic constraints.  In general, this is a manual process, since not all constraints are explicitly present in the code.
The main states of the process are as follows.
The first step is to turn type labels into channel labels.  We conjecture that, in many cases, we may be able to use names of variables in the application code that receives the data items.  If the channel value is used in semantic constraints that involve comparison of multiple values, we add an action that saves the received value in a variable of the EFSM.  We also add a guard that is a conjunction of predicates representing semantic constraints that use the channel value.  If the predicates involve other values, they are taken from the respective variables of the EFSM.

\begin{figure}
\centerline{\includegraphics[width=.8\textwidth]{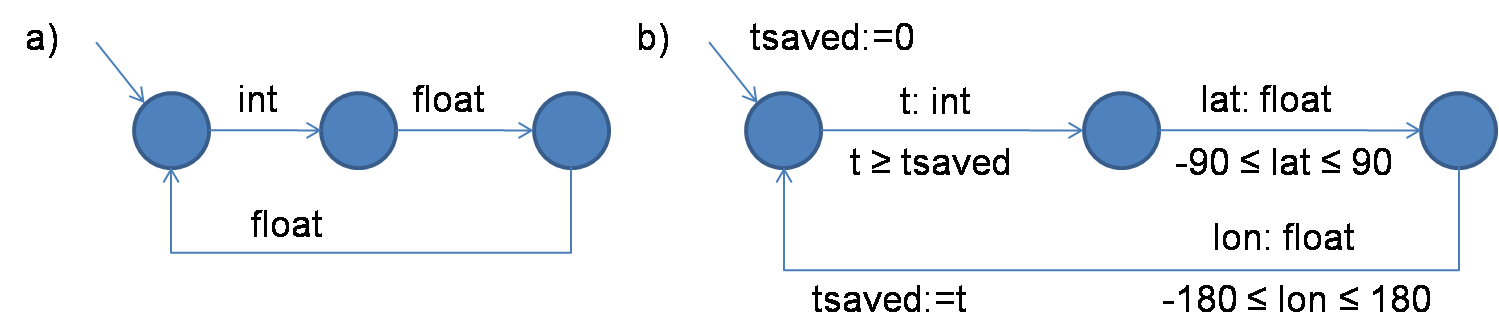}}
\caption{Extracting internal assumption from code: (a) state machine skeleton (b) EFSM with semantic constraints}
\label{fig:monitor}
\end{figure}

Figure~\ref{fig:monitor}, a) shows the skeleton extracted from the code of the case study from Listing~\ref{lst:simple_parser}, while Figure~\ref{fig:monitor}, b) shows the internal assumption EFSM after adding semantic constraints of the format.  Note that the same EFSM would be constructed for the code in Listing~\ref{lst:simple_parser2}.

At run time, we use the internal assumption EFSM as constructed above to monitor inputs read by the component application code.  Monitoring can be performed using runtime verification tools such as~\cite{Eagle04,Java-MaC04,MOP-overview}.

Before monitoring, we still need to check that the internal assumption EFSM is correct with respect to the assumption of the component.  In the approach of~\cite{Driscoll11}, extracted representation of the input format is verified against a similarly extracted output format of the producer by checking language containment: the language of the producer automaton should be included in the language of the consumer automaton.  In our case, we perform similar verification of the extracted internal assumption against the EFSM representing the assumption of the contract.  Before we do this, we need to modify the internal assumption EFSM to account for the semantics of I/O API calls.

\paragraph{Modeling I/O API effects.}
We extend the approach of~\cite{Driscoll11} with the ability to represent effects of I/O APIs used by the application code.  
An important effect of an I/O API is that it can make some aspects of the input format unobservable.  Here, we consider two kinds of I/O API and compare their effects.  First, consider the \texttt{scanf()} call used by the code in Listing~\ref{lst:simple_parser}.  Each call is parameterized by a format string, which is a sequence $\langle t_1, t_2,...\rangle$ of types.  The procedure skips over white space, including newline characters, preceding each item in the format string.  To represent this effect, we add a self-loop transition labeled by the \texttt{newline} symbol\footnote{Recall from Section~\ref{sec:case-study} that we are representing line breaks in the input format but not other kinds of white space.} to each state that occurs within the sequence.

\begin{figure}
\centerline{\includegraphics[width=.7\textwidth]{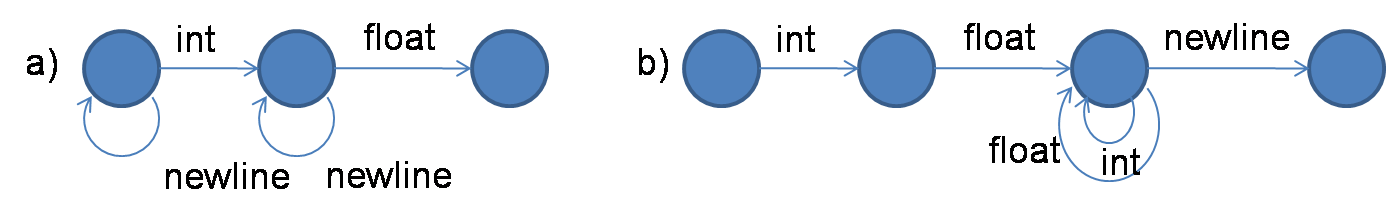}}
\caption{Extended skeleton fragments for (a) \texttt{scanf} call and (b) \texttt{readline/sscanf} pattern}
\label{fig:comparison}
\end{figure}

Now consider a different way to input data into the application code.  Input is first read into a string using the \texttt{getline} call, which is then processed using the \texttt{sscanf} call.  For the purpose of this paper, we consider this pattern as a single API call.  In this case, we know that the white space skipped by the call does not involve any line breaks, thus, there is no need to add the self-loop transition before each item in the format string.  However, any data item in the string beyond those listed in the format string is not read.  However, we know that the line break has been read when the call returns\footnote{For simplicity, we do not consider the end-of-file case here.}.  Thus the  sequence of transitions corresponding to a format string is extended by a self-loop transition for each non-newline symbol, and the sequence is extended with the newline-labeled transition at the end.  Figure~\ref{fig:comparison} shows extended skeleton fragments corresponding to the ``\%d\%f'' format string for the two cases.

Another important effect of the I/O API is type conversion performed by I/O routines.  For example, ``\%f'' directive will read an integer and convert it to the \texttt{float} type.  To capture this effect, we can make more complex skeleton fragments, containing alternative transitions.  We do not model type conversions in this paper. 

\paragraph{Monitor verification.}
In order to obtain the EFSM of the internal assumption that incorporates effects of the I/O API we repeat the construction outlined in the algorithm above.  Except now, in the skeleton extraction algorithm outlined above, instead of the skeleton fragment of the call, we use the extended fragments as described above.

Now we can check language inclusion between the two EFSMs.  This would ensure that no input that satisfies the assumption of the component will trigger a violation of the internal assumption of the code.  Internal assumptions for both versions of the calculator are correct in this respect.

\section{Discussion}
\label{sec:discussion}

\paragraph{Monitor construction.}
In this paper, we explored an approach to extract monitor skeletons from the application code using static analysis.  The extraction process, which inserts a state machine fragment for every I/O API call, allows us to incorporate API effects for monitor verification with respect to the assumption, by inserting a different state machine fragment.  The extraction process also identifies code locations that need to be instrumented in order to perform monitoring at run time.  Note that this process differs from most approaches to runtime verification~\cite{Eagle04,Java-MaC04,MOP-overview}, where monitors are developed independently of the code, based on requirements.  It is still possible, of course, to construct monitors independently.  The modification of the monitor to incorporate API effects for monitor verification would have to be done manually, too.

\paragraph{Detection power.}
What we need to consider is the detection power of monitors described above.  The first point to notice is that, since monitors are extracted from the code, there can be no structural violations.  By this we mean the case when an event $e$ is observed, but the monitor state machine does not expect $e$ to happen in the current state.  Thus, the monitor can detect only violations of semantic constraints. If constraints are too weak, deviations will go undetected.

As mentioned above, monitors do not observe all deviations from the input format specification.  However, the application code may be robust to some of the changes, effectively shielded by the choice of the I/O API.  Other changes lead to incorrect computations.

To illustrate this with our case study, consider two deviations from the input format specification.  One is to introduce a new field in each point.  This may correspond to a case where the sensor in the system is upgraded, and now produces altitude readings in addition to latitude and longitude.  We refer to this deviation as D1.  Another deviation, D2, is the removal of line breaks, which makes sense in a streaming context.  The first implementation of the calculator is robust to D2, and D1 is promptly detected by the monitor: the code attempts to read the altitude field as the next timestamp and then the timestamp as the latitude value, and one of the associated semantic constraints is likely to be violated.  Thus, not only is the error detected, but we get a stream of alarms that indicates that it is not an isolated format glitch, but a systematic problem.  The second implementation, on the other hand, is robust to D1: the whole point is read in, but the altitude field is not part of the format string and is never processed.  However, D2 is not detected, all points are read together as a single line, and only the first point is input.  No semantic constraints are violated, but the calculator produces a wrong result.  With stronger semantic constraints, this deviation could be detected.  For example, the input format could contain a field at the end of the stream, specifying the number of points in the input.  We would be able to check, then, that no points have been lost.

Ideally, we would like to design monitors that detect all deviations that make the code produce a wrong result and do not raise alarms when the component is robust to an observed deviation.  To achieve this, we need to characterize (1) which deviations can be detected and which cannot be, and (2) which deviations are harmful and which are not.  
To address question (1), we can look at the language difference between the two EFSMs.  That is, compute the set of runs that are rejected by the component assumption, but are accepted by the internal assumption of the application code.  These runs correspond to deviations from the format that are not detected.  The answer to question (2) seems harder.

Finally, we note that an alternative to instrumenting application code would be to instrument the I/O library, which we assumed to be a black box in this paper.  Such instrumentation would allow the monitor to have an unmediated view of input data, allowing us to check the input format specification directly.  On the other hand, instrumentation would have to be done at a much lower level of abstraction and require intimate understanding of how the library works.

\section{Conclusion}

We presented an approach for combining static, pre-deployment verification of assume-guarantee contracts for software components with dynamic, post-deployment monitoring of the contract assumption.  We identified a semantic gap between the contract assumption and the internal assumption of the application code of the component.  Unlike the contract assumption, which cannot be monitored directly, the internal assumption can be monitored by instrumenting application code.  We showed that, due to the semantic gap, not all violations of the assumption can be detected.  Our future work will concentrate on automated characterization of the semantic gap for a given contract assumption and I/O API used by the application code, in order to identify which critical deviations from the contract assumption can be missed by the monitor.

\bibliographystyle{eptcs}
\bibliography{references}

\end{document}